
\input phyzzx
\PHYSREV

\def\wdhat{\mathaccent"0362}
\def\wdtil{\mathaccent"0365}

\def\br{\overline}

\def\lag{{\cal L}}

\def\hlag{\wdhat{\cal L}}
\def\detg{\sqrt{-\rm{g}}}
\def\Pk{{\Phi_{\rm k}}}
\def\lym{{\cal L}_{\rm EYM}}
\def\leymh{{\cal L}_{\rm EYMH}}
\def\lpro{{\cal L}_{\rm ENAP}}
\def\a2{\left| A\right|^{2} }
\def\f2{\left| F\right|^{2} }
\def\delp{\left| \partial\phi\right|^{2} }
\def\dele{\left| \partial\eta\right|^{2} }
\def\om{\Omega}
\def\eijk{\epsilon_{ijk}}
\def\eabc{\epsilon_{abc}}
\def\fmni{ F^{(i)}_{\mu\nu}}
\def\fimn{ F^{(i)\mu\nu}}
\def\ami{A^{(i)}_{\mu}}
\def\ani{A^{(i)}_{\nu}}
\def\amj{A^{(j)}_{\mu}}

\def\ank{A^{(k)}_{\nu}}
\def\Vp{V\left( \phi\right)}
\def\Vpp{V^{\prime}\left( \phi\right)}
\def\VP{V\left( \Phi\right)}
\def\rr{R\left( r\right)}
\def\tr{T\left( r\right)}
\def\mr{m\left( r\right)}
\def\R2{\left(1-{2m\over r} \right)}
\def\drr{-\ln\left( R/ T\right)}
\def\dr{\delta\left( r\right)}
\def\edr{e^{-2\delta}}
\def\brr{\left( r\right)}
\def\bgrr{\left( gr\right)}
\def\brh{\left( r_{h}\right)}
\def\br0{\left( 0\right)}
\def\brinf{\left( \infty\right)}
\def\dom{\left(d\theta^{2} +\sin^{2}\theta d\varphi^2\right)}
\def\dx{d^4 x}
\def\pd{\partial}
\def\b2{b_{\mu}b^{\mu}}
\def\tar{{\wdhat \tau}_{r}}
\def\tat{{\wdhat \tau}_{\theta}}
\def\tap{{\wdhat \tau}_{\varphi}}
\def\taa{{\wdhat \tau}_{a}}
\def\tab{{\wdhat \tau}_{b}}
\def\tac{{\wdhat \tau}_{c}}

\def\c1{\left( 1+c\right)}
\def\w1{\left( 1+w\right)}
\def\ww1{\left( 1-w^2\right)}
\def\brt{\beta\left( r, t\right)}

\def\rwt{dr\wedge d\theta}
\def\rwp{dr\wedge d\varphi}
\def\twp{d\theta\wedge d\varphi}

\def\wp2{\left( w^{\prime}\right)^2}
\def\wwp{ w^{\prime}}

\def\gint{\int_{\om} \dx\detg}

\def\rhat{\wdhat r}
\def\What{\wdhat W}
\def\muhat{\wdhat \mu}
\def\lambhat{\wdhat \lambda}
\def\mhat{\wdhat m}
\def\rrh{\left( r-r_{h}\right)}
\def\wwrh1{\left[ 1-w^2\brh\right]}
\def\wrh1{\left[ 1+w\brh\right]}
\def\bbx{\left( x\right)}
\def\etanp{\left(\eta_{o}^{\prime}\right)}
\def\hofr{\left( {h\over r}\right)}
\def\ff{\left( 1+f\right)}

\magnification=1000
\chapterstyle={\Roman}
\FRONTPAGE
{\singlespace\centerline{
{}~~\hfil~~~~\hfil~~~~\hfil~~~~\hfil~~~~\hfil~~~~\hfil~~
{}~~\hfil~~~~\hfil~~~~\hfil~~~~\hfil~~~~\hfil~~~~\hfil~~
CLNS--92/1162~}}
{\singlespace\centerline{
{}~~\hfil~~~~\hfil~~~~\hfil~~~~\hfil~~~~\hfil~~~~\hfil~~
{}~~\hfil~~~~\hfil~~~~\hfil~~~~\hfil~~~~\hfil~~~~\hfil~~
{}~~~CTP--2160~}}
{\singlespace\centerline{
{}~~\hfil~~~~\hfil~~~~\hfil~~~~\hfil~~~~\hfil~~~~\hfil~~
{}~~\hfil~~~~\hfil~~~~\hfil~~~~\hfil~~~~\hfil~~~~\hfil~~
September 1992}}
\vskip 1.1in
{\singlespace\centerline{\fourteenpoint {\bf Eluding the No-Hair
Conjecture:}}
\centerline{\fourteenpoint {\bf Black Holes in Spontaneously Broken Gauge
 Theories}}}
\centerline{ }

{\singlespace\centerline{ }
\centerline{\twelvepoint  Brian R. Greene$^{\dagger}$
, Samir D. Mathur$^{\flat}$
and Christopher M. O'Neill$^{\dagger}$
}
\centerline{ }
 \centerline{\twelvepoint $^{\dagger}$Laboratory of Nuclear Studies}
\centerline{\twelvepoint Cornell University}
\centerline{\twelvepoint Ithaca, NY  14853 }
\centerline{ }
\centerline{\twelvepoint $^{\flat}$Center for Theoretical Physics}
\centerline{\twelvepoint Massachusetts Institute of Technology}
\centerline{\twelvepoint Cambridge, MA 02139} }

\normalspace
\centerline{ }
\centerline{ }
\centerline{ }
\centerline{\fourteenpoint Abstract}

We study regular and black hole solutions to the coupled
classical Einstein--Yang-Mills--Higgs
system. It has long been thought that black hole solutions
in the spontaneously broken phase of such a theory
could have no  nontrivial field structure outside of the horizon.
We first show that the standard  black hole
no-hair
theorem underlying this belief, although true in the abelian setting,
does not necessarily extend to the non-abelian case. This indicates
the possibility of solutions with non-trivial
gauge and Higgs configurations decaying exponentially {\it outside}
the horizon.
We then find such solutions by numerical integration of
the classical equations for the case of $SU(2)$ coupled to a Higgs doublet
(the standard model less hypercharge).
 As a prelude to this work we also study regular and black hole solutions
to Einstein--Non-Abelian--Proca theory and as a postscript we briefly discuss
the important issue of stability.
\vfill\supereject
{\parindent=0pt\bf\chapter{Introduction~~\hfil~~}}

Some time ago it was proven that black hole solutions in Einstein gravity
coupled to the abelian Proca model or to
a spontaneously broken abelian gauge theory cannot
have any nontrivial field structure outside of the horizon
\REF\bek{J.D.~Bekenstein\journal Phys.Rev.&D~5(1972)1239.}
 \REF\adl{S.L.~Adler and R.D.~Pearson\journal Phys.Rev.&D~18(1978)2798.}
\REF\teit{C.~Teitelboim\journal Phys.Rev.&D~5(1972)2941.}
\REF\price{R.H.~Price\journal Phys.Rev.&D~5(1972)2419.}
[\bek --\adl].
 In conjunction
with similar arguments applied to a variety of field theories coupled to
gravity   [\bek --\price],
 a widely held belief has been that
the only distinguishing features of a black hole exterior
to the horizon are charges carried by massless gauge fields
\REF\colea{S.~Coleman, J.~Preskill, and F.~Wilczek
\journal Nucl.Phys.&B~378(1992)175.}
\REF\foota{In this paper our concern is with classical ``hair'' only.
We do not address the interesting but separate issue of
quantum hair
 [\colea ].} [\foota ]. These are
mass, angular momentum and ``electric'' or ``magnetic'' charges, where the
latter refer to charges carried by  some
possibly non-abelian gauge group. This belief is often referred to as the
black hole no-hair theorem (or more precisely, no-hair conjecture).

In elementary particle physics we are familiar with the fact that
{\it non-abelian} symmetries often play a crucial role.
Based upon the abelian case and more general prejudice, it has
 generally been accepted  that
the no-hair statements  ensure that no evidence of the
broken classical charges can be found outside a black hole horizon.
In particular, it has been thought that one cannot find stationary
black hole solutions which are nonsingular at the event horizon and have
a nonvanishing massive vector field that decays exponentially outside of the
horizon
\REF\coleb{S.~Coleman, J.~Preskill, and F.~Wilczek\journal
Phys.Rev.Lett.&67(1991)1975}
\REF\gibbons{G.W.~Gibbons,
``Self-gravitating Magnetic Monopoles, Global Monopoles and
Black Holes'', {\it Lisbon 1990, Proceedings, The Physical Universe}
(1990)110.}
(for example, see [\bek --\adl],[\colea] and [\coleb --\gibbons]).
It is the intent of the present paper to quantitatively
investigate the veracity of this belief. That is, we
study black hole solutions in Einstein
gravity coupled to the non-abelian Proca model and to a spontaneously broken
gauge theory (with symmetry group $SU(2)$ in each case) in order to
determine whether the no-hair statements are affected by the non-abelian
structure. Our results indicate that, in fact, the non-abelian nature of the
theory has a crucial impact giving rise to classical solutions which
violate the no-hair conjecture.

In particular, we numerically study spherically symmetric solutions to
Einstein gravity coupled to $SU(2)$ gauge theory and a Higgs doublet,
as in the standard model (without hypercharge).
We take the standard model form for the symmetry breaking Higgs potential
with only the lower real component of the Higgs field having a nonzero
vacuum expectation value.
 Integrating via a two parameter shoot
(associated with the initial values of the gauge and Higgs fields), we find
black hole solutions in which there are nontrivial gauge and Higgs fields
outside of the horizon  (with no global charges associated with the
gauge or Higgs fields)
that ultimately exponentially decay to their
vacuum values. In fact, we find two families of such solutions with each
member distinguished by the number  $k$ of zero crossings of the single free
function in the gauge connection. The $k=1$ member of one of these two
families of solutions has a limit (the dimensionless Higgs vacuum expectation
value going to zero) in which gravity becomes arbitrarily weak
outside of the horizon and the solution  resembles the known flat space
\REF\sphnote{ References to sphaleron solutions in this paper denote the
standard SU$(2)$ YM solution with a Higgs doublet (rather than more
general saddle points).}
sphaleron solution [\sphnote]. So, at least this particular solution might be
characterized as a being a black hole with sphaleron hair. Our study indicates,
however, that the other solutions do not have a weak gravity limit
and hence do not have known flat space counterparts.

We note that ours is not the first challenge to the no-hair conjecture, but it
is useful and important to emphasize the viewpoint espoused in [\colea].
As these authors point out, one should
distinguish primary hair
from secondary hair. The latter refers to black hole structures which
exist solely as the result of (well known)
primary hair such as gauge charges
and hence are not fundamentally new characteristics. The existence
of secondary hair,
therefore, is not really in conflict with the no--hair conjecture.
The
charged dilaton black hole solutions
\REF\strom{G.W.~Gibbons and K.~Maeda\journal
Nucl.Phys.&B~298(1988)741.}
\REF\gar{D.~Garfinkle, G.T.~Horowitz and A.~Strominger\journal
Phys.Rev.&D~43(1991)3140.}
\REF\press{J.~Preskill, P.~Schwarz, A.~Shapere, S.~Trivedi, F.~Wilczek\journal
Mod.Phys.Lett.&A~6(1991)2353.}
\REF\shap{A.~Shapere, S.~Trivedi and F.~Wilczek\journal
Mod.Phys.Lett.&A~6(1991)2677.}
\REF\holz{C.~Holzey and F.~Wilczek\journal Nucl.Phys&B~380(1992)447.}
\REF\horne{J.H.~Horne and G.T.~Horowitz\journal
Phys.Rev.&D~46(1992)1340.}
[\strom --\horne ] in which there is
 nontrivial scalar field configuration outside of
the horizon provides an example of this distinction.
The dilaton
configuration is nontrivial because the electric charge (primary hair) acts
as a source and hence yields only secondary hair. Another interesting
solution which (among other things) provides a  challenge to
the no-hair theorems are the ``black holes inside magnetic monopoles''
found in
\REF\leenair{K.~Lee, V.P.~Nair, and E.J.~Weinberg
\journal Phys.Rev.&D~45(1992)2751.}
\REF\ortiz{M.~Ortiz
\journal Phys.Rev.&D~45(1992)2586.}
[\leenair, \ortiz ]. Again, however, the hair associated with the nontrivial
field configuration outside of the horizon is of the secondary sort
[\colea ].
 \REF\biz{P.~Bizon\journal Phys.Rev.Lett.&64(1990)2844.}
\REF\bart{R.~Bartnik and J.~McKinnon\journal Phys.Rev.Lett.&61(1988)141.}
Two other challenges
to the no-hair conjectures, which are closer to having
primary hair, are
the works of  [\biz ] (motivated by
the paper of Bartnik and McKinnon [\bart ]) and
\REF\strauc{S.~Droz, M.~Heusler, and N.~Straumann\journal
Phys.Lett.&B~268(1991)371.}
[\strauc ]. The work of [\biz ] finds numerical
 black hole solutions of Einstein gravity
coupled to $SU(2)$ gauge theory in which the fields decay sufficiently
quickly so as to have vanishing global charges. The hair on these solutions
is still carried by massless gauge fields -- the novel feature is that
the fields leave no imprint
at infinity. Unfortunately, these solutions have been shown to be unstable
\REF\straub{Z.~Zhou and N.~Straumann\journal Nucl.Phys.&B~369(1991)180.}
 [\straub ].  In [\strauc ]
some interesting numerical work also has
 shown that black holes can exist with nontrivial Skyrme fields in
the Einstein--Skyrme model.

Our work on this subject was partly
motivated by the publication of [\bart ] in which
the authors found smooth solutions to the coupled Einstein--$SU(2)$ gauge
theory classical equations
\REF\yau{J.~Smoller, A.~Wasserman, S.-T.~Yau, and J.~McLeod
\journal Commun.Math.Phys.&143(1991)115.}
\REF\footc{In [\bart] strong numerical evidence was given for the
existence of such solutions. Subsequently,
the existence of these solutions was rigorously established in [\yau].}
[\footc].
These smooth solutions were surprising in that
no-go theorems for classical glueball solutions to non-abelian gauge theory
(without coupling to gravity)
were established in \REF\colec{S.~Coleman\journal Commun.Math.Phys.&55(1977)
113.}
\REF\deser{S.~Deser\journal Class.Quantum Grav.&1(1984)L1.}
[\colec]
while no-go theorems for smooth solutions including
gravity were proven for $2 + 1$ dimensions in [\deser].
Unfortunately, the $ 3 + 1 $ dimensional solutions of [\bart ] were
subsequently proven to be unstable.
A natural question to ask
is whether such smooth solutions in $3 + 1 $ dimensions can be found
in the physically relevant case of spontaneously broken non-abelian gauge
theories and, if so, whether they might be stable. In addition to our black
hole solutions, we do in fact find  and present
such smooth solutions but have not as
yet exhaustively analyzed their stability properties.

Stability is, in fact,
an important unanswered question about both our black hole and regular
solutions. However, we hasten to
emphasize that the long held belief that such black hole
solutions do not exist is based on
the no-hair results which themselves have nothing to do with stability.
Rather, the no-hair
results were based on the careful examination of  certain
classical field theories coupled to gravity
which simply have a different character from
the non-abelian vector theories studied here. However, as far as physical
relevance is concerned, stability is a crucial feature.
General prejudice would certainly lean towards suspecting that our
solutions are unstable for two main reasons. First, without a global charge
leaving an imprint at infinity, stability certainly seems less likely. Second,
as mentioned, there is a relation between our solutions and flat space
sphalerons, the latter of which are unstable.
On the other hand, some of our solutions seem to involve the trapping of gauge
bosons in a region of space where, due to the Higgs configuration, they are
less
massive (relative to their asymptotic mass), and this, potentially, may bode
well for stability.
 It has been noted that
there might be a no--hair theorem if one only considers stable solutions
[\straub ]. We do not know, however,
 of any definitive arguments supporting this
natural suggestion.
 We are presently studying the stability question for both
the smooth and black hole solutions
discussed here and will report on this elsewhere.

The organization of this paper is as follows. In section II  we review the
no-hair theorems and indicate why they do not apply in the situation under
study. This opens up the possibility for the existence of black hole
solutions with non-abelian gauge and Higgs hair. In section III we
embark on the numerical construction of such solutions, beginning with the
simpler Einstein--Non-Abelian--Proca system for the case of $SU(2)$.
This system is interesting in its
own right as a first concrete illustration of how the
classical solitons in abelian and
non-abelian Proca theories differ substantially. It also is a useful prelude
to our discussion of the Einstein--Yang-Mills--Higgs system in section IV.
In both sections III and IV we present spherically symmetric regular and
black hole solutions. In section V we present some preliminary work
on the stability and we offer our conclusions.

{\parindent=0pt\bf\chapter{ Limitations on No-Hair Theorems~~\hfil~~}}
\let\chapterlabel=2

Many of the no-hair proofs for classical field theories utilize an
elegant and powerful method first developed by Bekenstein [\bek].
It is, in effect, a repackaging of the generally covariant
field equations into a statement about the behavior of fields
outside the event horizon. In this section, we briefly review
Bekenstein's arguments and illustrate how non-abelian gauge
fields in general, and the Einstein--Non-Abelian--Proca (ENAP)
and Einstein--Yang-Mills--Higgs (EYMH) systems in particular,
can possess static black hole solutions with nontrivial exterior
structure.

Consider the action for a set of arbitrary local
fields $\Pk$ in a gravitational background: $S= \int\dx
{\hlag}\equiv\int\dx\detg\lag$.
After multiplication by $\dx\Pk$ and integration by parts, the covariant
Euler-Lagrange equations become
$$
\sum_{\rm k}\int_{\om}\dx
\pd_{\mu}\left[\Pk {\pd\hlag\over \pd\left(\Pk_{,\mu}\right)}\right]
=\sum_{\rm k}\int_{\om}\dx\left[ \Pk_{,\mu}
{\pd\hlag\over \pd\left(\Pk_{,\mu}\right)} + \Pk{\pd\hlag\over \pd\Pk}
\right].\eqn\bekint
$$
The left-hand side of this equation can be expressed as a surface
integral $\int_{\pd\om}dS_{\mu}b^{\mu}$, where $dS_{\mu}$ is an
element of the hypersurface bounding the volume $\om$ over which we integrate,
and
$$
b^{\mu}\equiv\sum_{\rm k} \Pk {\pd\hlag\over \pd\left(\Pk_{,\mu}\right)}.
\eqn\bmudef$$
If we assume that our
system admits black hole solutions and we choose our four-volume to be the
black hole exterior, then $\pd\om$ consists of
the horizon, spatial infinity, and
future and past timelike infinities. The behavior of $b^{\mu}$ at spatial
infinity is already determined for all nontrivial and physically relevant
fields by the field equations: $b^{\mu}\rightarrow 1/r^3$
asymptotically for massless fields and $b^{\mu}$ vanishes exponentially
for massive fields, so there is no contribution at spatial infinity
to the left side of \bekint. Since $dS_{\mu}$ is proportional to the normal
$n_{\mu}$ of the hypersurface, which can be chosen to satisfy
$n_{i} = 0$ as
$t\rightarrow\infty$, $\ \ dS_{\mu}b^{\mu}$ also vanishes at timelike
infinity when $b^{0}=0$. This condition on $b^{0}$  is satisfied by
static fields, and thus there are no non-horizon contributions to the surface
integral in eq.\bekint . From the fact that the horizon is a null hypersurface
$\left( g_{ij}dS^i dS^j = 0\right)$ and that $g_{ij}$ is positive semi-definite
on the horizon, it can easily be shown that $dS_{\mu}b^{\mu}$ vanishes
on the horizon when $b_{\mu}b^{\mu}$ is bounded there. Thus, for static fields
with finite $b_{\mu}b^{\mu}$ on the horizon, eq.\bekint
 ~implies that the integral over the entire black hole exterior
of some nontrivial function of the fields and their derivatives  must vanish.
The strategy for establishing no hair becomes clear: if one can demonstrate
that the integrand in eq.\bekint ~is negative or positive definite,
 then the only finite energy solutions
 satisfying \bekint
 ~are those for which the integrand vanishes.
For most Lagrangians of physical interest,
this implies that the fields must assume
constant values over the entire black hole exterior.

The Bekenstein approach has been used to establish no-hair theorems for
Klein-Gordon theory, abelian Proca [\bek] and Higgs theories,
 and the Goldstone model
[\adl ], among others.
Here we provide a sketch of the right-hand side of
eqn.\bekint ~for non-abelian gauge fields
which demonstrates the possibility of the existence of
 Einstein--Yang-Mills (EYM) black holes.

The Lagrangian for an $SU(2)$ gauge theory may be written
$$
\lym = -{1\over 4\pi}
\left( {1\over 4}\f2\right)\equiv
 -{1\over 4\pi}\left( {1\over 4}g^{\mu\rho}g^{\nu\sigma} F^{(i)}_{\mu\nu}
F^{(i)}_{\rho\sigma}\right) ,\eqn\elyma
$$
where $i$ is the isospin index. A straightforward calculation of the
right side of eq.\bekint ~for a metric
with signature $(-1,+1,+1,+1)$ gives
$$
 -{1\over 4\pi}\int_{\om}\dx\detg\left(
-8\pi\lym +{1\over 2}\left[g\eijk\amj\ank\right]\fimn\right),\eqn\bekym
$$
where $g$ is the gauge coupling: $ \fmni =\pd_{\mu}\ani -
\pd_{\nu}\ami +g\eijk\amj\ank $. If we consider only static fields, and
for simplicity assume $A_{0}^{(i)}=0$, then
$\f2\ge 0$. The second term in
the integrand, however, is not necessarily positive: its sign will depend
on the details of the solution in the black hole exterior $\om$, so it is
clear that the nonlinear terms can lead to a possible way of avoiding this
no-hair argument.

 To demonstrate this point more
quantitatively,  consider the
expression in \bekym. Let $C_{\mu ,\nu}^{(i)}=\partial_\mu
A^{(i)}_\nu - \partial_\nu A^{(i)}_\mu$ and $D_{\mu ,
\nu}^{(i)}=g\epsilon_{jk}^{i} A_\mu^{(j)}A_\nu^{(k)}$. Then to avoid  the
Bekenstein argument  we require the non-positivity (somewhere) of
$S\equiv (C+D)\cdot (C+D)+D\cdot (C+D)$
 where $C\cdot C=C_{\mu ,\nu}^{(i)} C^{(i)\mu \nu}$ etc.
 With the gauge choice $A_0^{(i)}=0$,  and the
 assumption that in this gauge all fields are time
 independent (no `electric fields'),
 it follows that $(\alpha
C+\beta D)\cdot (\alpha C+\beta D)\geq 0$ for any real $\alpha, \beta$.
The relation $S=(C+2D)\cdot (C+D)\leq 0$ requires that $C\cdot D\leq 0$,
and gives $C\cdot C+2D\cdot D\leq -3C\cdot D$. This may be
rephrased as
 $${1\over 2}(C\cdot C+D\cdot D)-{1\over 6}(C\cdot C-D\cdot D)~\leq
{}~-C\cdot D~\leq~{1\over 2}(C\cdot C+D\cdot D) ,\eqn\inequal
$$
where the second inequality is just the positivity of $(C+D)\cdot
(C+D)$.

Noting that $D$ is the non-abelian term we see that the non-abelian
nature of the field is crucial in avoiding the Bekenstein argument
for no-hair. The above relation also shows that the value of $C\cdot D$
must lie in  a very restricted range for the non-positivity of $S$: the
cubic term in the fields involving the structure constants is bounded
both above and below by the quadratic and quartic terms in the gauge
fields.
(It is possible to make restricted gauge transformations
that preserve time independence of fields
and  $A_0^{(i)}=0$. What we see is that \inequal ~ must
hold in some region in each such gauge.)

The left-hand side
of eq.\bekint ~also presents complications. For a gauge field, $\b2$
is not a physical scalar and need not be bounded on the horizon, so
 a practical application of the Bekenstein approach requires the explicit
evaluation of
$$
b^{\mu}=-{1\over 4\pi}\fimn\ani \eqn\bmuym
$$
on the horizon to make the sign considerations for eq.\bekym ~relevant.
By introducing the {\it ansatz} for the connection $A$ used in
[\bart ] and [\biz ],
 it can be shown that $\b2$ is bounded on the horizon and that
 the second term in \bekym ~can be negative in the black hole exterior, so
that a black hole with nontrivial gauge field structure outside the horizon is
allowable (but not necessary) under the Bekenstein analysis. In fact,
the solutions of [\biz ] are examples of these ``colored'' black holes.

When this approach is applied to the ENAP theory considered below, the results
are similar. For our metric convention, the Lagrangian is
$$
\lpro =  -{1\over 4\pi}
\left( {1\over 4}\f2 + {\mu^2\over 2}\a2\right) ,\eqn\elpro
$$
with $\a2\equiv g^{\mu\nu}A^{(i)}_{\mu}A^{(i)}_{\nu}$ and $\mu$ is the vector
field mass. As we stated above, this is obviously not a gauge-invariant
theory but provides a good first approximation for a massive non-abelian
gauge field. For the massive field, $b^{\mu}b_{\mu}$
 is now a physical scalar bounded on the horizon, and
the Bekenstein integral \bekint ~assumes the same form as in \bekym, with
$\lym$ replaced by $\lpro$. For $A^{(i)}_{0}=0$, the same arguments
as above establish the possibility of hair for  massive vector
black hole solutions.

The second theory we investigate in this paper is EYMH, whose
Lagrangian is given by
$$
\leymh =\lym -{1\over 4\pi}\left[ \left( D_{\mu}\Phi\right)^{\dagger}
\left( D^{\mu}\Phi\right) +\VP\right] ,\eqn\eleymha
$$
where $D_{\mu} = \pd_{\mu} +g\  {\bf \tau}\cdot {\bf A}_{\mu}$ is the usual
gauge-covariant derivative expressed in the antihermitian representation
of $su(2)\  $ ( $ \tau_{i}= -i\sigma_{i} / 2$). The Higgs field $\Phi$ is
taken to be a complex doublet $\Phi = (\phi^{+},\phi^{-} )$ and $\VP $
to be a double-well potential with degenerate vacua.
 Though we give further details
of our Higgs {\it ansatz} below, for now
 we assume $\Phi$ possesses only one degree of freedom:
$$
\Phi ={1\over \sqrt{2}}\left(\matrix{
0 \cr
\phi\left( {\bf x}\right) \cr}\right) ,\eqn\phidef
$$
 for $\phi $ real and time-independent. The Lagrangian \eleymha ~then becomes
$$
\leymh =  -{1\over 4\pi}
\left( {1\over 4}\f2 + {1\over 2}\left({g\phi\over 2}\right)^2\a2
+{1\over 2}\delp +\Vp\right) , \eqn\eleymhb
$$
where $\delp\equiv g^{\mu\nu}\pd_{\mu}\phi\pd_{\nu}\phi\ge 0$,
 and $(g\phi / 2)$ now
plays the role of the gauge field mass. The vector $b^{\mu}$ has an
additional term
$$
b^{\mu}=-{1\over 4\pi}\left[ \fimn\ani +g^{\mu\nu}\left(\pd_{\nu}\phi\right)
\phi\right] ,\eqn\bmueymh
$$
and the Bekenstein integral is
$$\eqalign{
-{1\over 4\pi}\int_{\om}\dx\detg  &\left[
{1\over 2}\f2 +\left({g\phi\over 2}\right)^2\a2
+{1\over 2}\left[g\eijk\amj\ank\right]\fimn\right]  \cr
-{1\over 4\pi}&\int_{\om}\dx\detg  \left[
 \delp + \left({g\phi\over 2}\right)^2\a2
+\Vpp\phi\right],\cr}\eqn\bekeymha
$$
with $\Vpp\equiv d\Vp /d\phi $. The separate integrals in this
expression correspond to the YM
 and Higgs surface integrals obtained from eq.\bmueymh.
The YM
term is identical in form to the Proca case, so by the above arguments YM
hair is possible. In the Higgs term, the sign of $\Vpp\phi$ for a double-well
$\Vp$ obviously depends on the details of a solution, so the second integrand
need not be positive definite either. Black hole solutions to EYMH theory
having both nontrivial YM and Higgs field structure are therefore
not excluded by the Bekenstein analysis.

It is important to emphasize that these sign arguments only demonstrate
the {\it possibility} of solutions with hair. An example for which all of the
above reasoning is valid but a no-hair proof has been found [\adl ] is
the pure Higgs (Goldstone) theory, which corresponds to the Higgs contribution
to \bekeymha ~with $\a2 \equiv 0$. The Higgs-gauge field coupling
and the non-abelian gauge interactions in EYMH
theory therefore play a critical role in the existence of black hole solutions
for which $\phi\not\equiv constant$ outside the horizon. We should also
mention that Price [\price]  has employed somewhat different
reasoning to argue for a black hole no-hair theorem. His arguments, though,
are only valid in the domain of weak fields (in which the important
nonlinear terms would lose their influence) and hence do not directly apply
to the present context. In the following we shall show that the gap in the
no-hair arguments described is wide enough to allow for solutions which
have spontaneously broken gauge and Higgs hair.

{\parindent=0pt\bf\chapter{ ENAP Theory: Regular and Black Hole Solutions
{}~\hfil~~}}
\let\chapterlabel=3

{\parindent=0pt\bf Metric and Connection}

The metric for a static, spherically symmetric spacetime
can be written
$$
ds^2=-T^{-2}\left( r\right) dt^2 + R^{2}\left( r\right) dr^2 +r^2\dom
,\eqn\meta
$$
where $\rr\equiv\left( 1-2m/r\right)^{-1/2}$ and $\mr$ may be
interpreted as the total
mass-energy within the radius $r$. An alternate form of the metric convenient
for describing black hole solutions follows from defining $\delta\equiv\drr$.
In terms of the functions $\left( \dr ,\mr\right)$, the metric becomes
$$
ds^2=-\R2\edr dt^2 +\R2^{-1} dr^2 +r^2\dom .\eqn\metbh
$$
For the metric \meta, regularity at the origin requires
$ T\br0 < \infty$ and $R^{\prime}\brr , T^{\prime}\brr \rightarrow 0$ , while
asymptotic
flatness requires $\rr,\tr\rightarrow 1$. For the alternate form of the
metric, a regular event horizon at $ r=r_{h}$ requires
$$
m\brh = r_{h}/2, \ \ \ \ \ \delta\brh <\infty ,\eqn\bhbc
$$
and comparison with the asymptotics of eq.\metbh ~gives $\dr\rightarrow 0$
as $r\rightarrow \infty$.
 For static solutions, however, we can rescale the time coordinates
$$\eqalign{
d{\wdtil t} &\equiv T^{-1}\br0 dt \cr
d{\wdtil t} &\equiv e^{-\delta\brh } dt \cr }\eqn\tdela
$$
in order to simplify the initial data problem: for such a rescaling,
$$
\eqalign{
{\wdtil T }\br0 &= 1, \ \ \ \ \ {\wdtil T }\brinf = 1/T_{o}\equiv {1/ T\br0
}\cr
{\wdtil\delta }\brh &= 0 , \ \ \ \ \ {\wdtil\delta }\brinf =-\delta_{o}\equiv
-\delta\br0 .\cr}\eqn\tdelb
$$
Thus the correct initial values of $g_{tt}$ for a given system can be
determined
by integrating the Einstein equations with the fully specified initial
conditions
$$
\eqalign{
R\br0 = 1, \ \ \ \ \ &T\br0 = 1\cr
 m\brh =r_{h}/2,\ \ \ \ \ &\delta\brh =0  \cr}\eqn\tdelc
$$
where in each of the last three equations, the top part refers to regular
solutions and the bottom part to black hole solutions.
We will use this parameterization below.

The most general spherically symmetric $SU(2)$ connection
 \REF\wit{E.~Witten\journal Phys.Rev.Lett.&38(1977)121.}[\wit] is
$$
A={1\over g}\left[ a\tar\ dt +b\tar\ dr +\left( d\tat -\c1
\tap\right) d\theta +\left( \c1\tat +d\tap\right) \sin\theta d\varphi\right],
\eqn\conna
$$
where $a$ and $b$ have dimensions $[L]^{-1}$, $c$ and $d$ are dimensionless,
$g$ is the gauge coupling and
$\left(\tar,\tat,\tap\right)$ is the
antihermitian $su(2)$ basis expressed in the usual
$3$-d (physical space) polar coordinate directions; e.g. $\tar=
{\bf\wdhat r}\cdot{\bf \tau},\ \ {\rm and}
\left[ \taa,\tab\right] =\eabc\tac $ with the indices ranging over
$\left( r,\theta,\varphi\right)$.
The four degrees of freedom $a,b,c,d$ are all functions of the $3$-d
radius $r$ and time $t$. The connection \conna ~has a residual gauge freedom
$$
A\longrightarrow UAU^{-1} + {1\over g}UdU^{-1}\eqn\gta
$$
under unitary transformations of the form $U=\exp\left[ \brt \tar\right]$,
where $\brt$ is an arbitrary real function.
Under such gauge transformations, which form an
abelian subgroup of the full gauge group, the connection functions
transform as
$$
U:\left( \matrix{
 a \cr b \cr c \cr d \cr}\right)
 \longmapsto\left( \matrix{
 {\wdtil a} \cr {\wdtil b} \cr {\wdtil c} \cr {\wdtil d} \cr}\right)
 =\left( \matrix{
a-{\dot\beta} \cr b-\beta^{\prime} \cr
c\cos\beta -d\sin\beta \cr
d\cos\beta +c\sin\beta \cr}\right) .\eqn\funtran
$$
We can use this freedom to impose the ``polar gauge'' ${\wdtil b}\equiv 0$ .
In the static case, ${\dot \beta}=0$ and ${\wdtil a},{\wdtil c},{\wdtil d}$
are functions of $r$ only. Following
[\bart] and [\biz], we eliminate two of the
three remaining degrees of freedom by the {\it  ansatz} ${\wdtil a}\equiv 0$
(no dyons, purely magnetic YM curvature) and ${\wdtil d}\equiv 0$.
\REF\straua{N.~Straumann and Z.~Zhou\journal Phys.Lett.&B~237(1990)353.}

After relabelling the remaining degree of freedom $w\brr\equiv {\wdtil c}\brr$,
the connection assumes the form
$$
A={1\over g}\w1\left[ -\tap d\theta +\tat\sin\theta d\varphi\right]\eqn\connb
$$
which was  explored by t'Hooft in the context of magnetic monopoles
\REF\thft{G.~t'Hooft\journal Nucl.Phys.&B~35(1971)167.} [\thft].
It differs from the  {\it ansatz} of [\bart] by a singular gauge transformation
$U=\exp\left(\theta \tau_{1}\right)\exp\left(\left({\pi\over 2}-
 \varphi\right)\tau_{3}\right) $ and has the added virtue
that $\a2 $ is invariant under spatial rotations.

{\parindent=0pt\bf  }
{\parindent=0pt\bf ENAP Equations and Boundary Conditions}

Defining $F =dA+ g A\wedge A$ , we have
$$
F={w^{\prime}\over g}\left[ -\tap \rwt +\tat\sin\theta\rwp\right]
-{1\over g}\ww1\tar\sin\theta\twp .\eqn\fyma
$$
The Proca equations $D{}^{\ast} F +\mu^2{}^{\ast}A= 0$
reduce to a single equation:
$$
{d\over dr}\left( {w^{\prime}\over RT}\right) +{w\ww1\over r^2}{R\over T}
-\mu^2\w1 {R\over T} =0.\eqn\ymprob
$$

The  Einstein equations $G_{\mu\nu}=8\pi T_{\mu\nu}$ may be calculated
by varying with respect to $g_{\mu\nu}$ the action
$$
\gint\left( {R\over 16\pi}+\lpro\right) ={1\over 16\pi}\gint
\left(  R-\f2 -2\mu^2\a2\right) ,\eqn\acpro
$$
where $R=R_{\mu\nu}g^{\mu\nu}$ is the gravitational curvature
 scalar and we have set $G=1$. The energy-momentum tensor is
$$
8\pi T_{\mu\nu} = 2F_{\mu\gamma}F_{\nu}{}^{\gamma} -{1\over 2}g_{\mu\nu}
\f2 +\left[
2\mu^2 A_{\mu} A_{\nu} -g_{\mu\nu}\mu^2\a2\right] ,\eqn\tabpro
$$
and the $(tt)$ and $(rr)$ Einstein equations can be written
$$\eqalignno{
m^{\prime} =&\R2\left( w^{\prime}\over g\right)^2 +{1\over 2}{\ww1^2\over
g^2r^2}
+{\mu^2\over g^2}\w1^2 &\eqname\meqpro\cr
r\R2{T^{\prime}\over T} =& -\R2\left( w^{\prime}\over g\right)^2 +
{1\over 2}{\ww1^2\over g^2r^2} &\eqname\teqpro\cr
 &\phantom{-\R2\left( w^{\prime}\right)^2 }\;+{\mu^2\over g^2}
\w1^2 -{m\over r}&\cr}
$$
in terms of the metric functions $\mr$ and $\tr$.
While \teqpro ~is appropriate for obtaining regular solutions, the
coordinate singularity at the event horizon of a black hole solution
leads us to replace it with
$$
\delta^{\prime} =-{2\left( w^{\prime}\right)^2\over g^2 r} ,\eqn\deqpro
$$
where $\delta $ was introduced in \metbh ~above. For either choice of
metric functions, the two Einstein equations can be used to express the Proca
equation \ymprob ~in a form independent of $\tr$ or $\dr$:
$$\eqalignno{
r^2\R2w^{\prime\prime}+\left[ 2m-{\ww1^2\over g^2 r}-2\ {\mu^2\over g^2}
\w1^2r\right]
&w^{\prime} &\eqname\ymproc\cr
+ \ww1 w -\mu^2&\w1 r^2 =0. &\cr}
$$
All of the above agrees with the corresponding EYM results [\bart ,\biz]
 when we
take $\mu =0$.

The presence of the additional parameter $\mu$ in ENAP theory motivates
us to clarify the role of the gauge coupling $g$ in the field equations.
A classical theory with $G=c=1$ satisfies $[L]=[T]=[M]$, which for our
theory implies $\left[ g\right]=\left[\mu\right]=\left[ L\right]^{-1}$.
If we introduce the dimensionless quantities
$$
{\wdhat r}\equiv gr, \>\>\>\>\>
{\mhat\left(\rhat\right)\equiv g\ m\left( gr\right) },\>\>\>\>\>
{\wdhat \mu}\equiv \mu /g ,\eqn\dimquant
$$
then the theory scales as
$$
\f2 = g^2{\left| F\left(\rhat\right)\right|}^{2}, \ \ \ \ \
\a2 ={\left| A\left(\rhat\right)\right|}^{2}, \ \ \ \ \
\lpro = g^2\lpro\left(\rhat ,\muhat \right) ,\eqn\fal
$$
and the dimensionless ENAP equations assume the form of
\meqpro --\ymproc ~with $g=1$ and
\dimquant ~replacing the dimensionful variables and parameters.
Through scaling, we may therefore obtain a solution to
 the ENAP equations for any $g>0$ from a solution to the dimensionless
equations:
$$\eqalign{
w_{g}\brr =\What\bgrr \ \ \ \ \ \  & R_{g}\brr=
\left( 1- {2\mhat\bgrr /\bgrr}\right)^{-1/2}\cr
m_{g}\brr ={1\over g}\mhat\bgrr  \ \ \ \ \ &T_{g}\brr=T\bgrr; \ \delta_{g}\brr
=\delta\bgrr .\cr}\eqn\scalesoln
$$
Solutions with $g>1$, for example, possess the same structure as $g=1$
 solutions, but it occurs at radius $r=\rhat /g<\rhat$.
For the remainder of this section, we take $g=1$ without loss of generality.

We can anticipate from  the field equations the
general characteristics of solutions.
{}From the definition of $\delta $, eq.\deqpro ~demonstrates that $R/T$
increases
monotonically with radius. $T$ can also be shown to satisfy
the same equation as in EYM theory [\bart],
$$
{d\over dr}\left[{r^2\over R}\left( {1\over T}\right)^{\prime}\right]
= 2\R2\wp2 {R\over T} +\left( {R\over r^2 T}\right)\ww1^2 ,\eqn\tconpro
$$
so that $T^{\prime}<0$ for $w\not\equiv 1$. From the boundary conditions of
the unrescaled metric ($T\br0 >R\br0 =1, \ T\brinf =R\brinf =1$) and the
definition of $R$, we then expect that $T>R\ge 1$, $R$ should possess at least
one maximum, and $T$ should decrease monotonically for nontrivial regular
solutions. For black hole solutions, $\delta$ should decrease monotonically
and $R$ should decrease from its singular value at the horizon, though it
need  not  possess any local extrema for $r_{h} < r <\infty $.

Following [\bart],
we can learn about the behavior of the vector field by rewriting the Proca
equation in the form
$$
{1\over 2}{d\over dr}\left[ {\left( w^2\right)^{\prime}\over RT}\right] =
{\wp2\over RT}+{R\over r^2 T}\left[\left(w^2-1\right) w^2+\mu^2\w1 wr^2\right]
{}.
\eqn\ymhair$$
A trivial $m=0$   solution occurs for $w\equiv -1$, and comparison with
eq.\meqpro ~re\-veals that $w=-1$ is the only acceptable asymptotic value for
finite energy solutions. Since the right-hand side of this equation is
manifestly positive for $w^2>1$, the only nontrivial
solutions  having finite $w^2$ and finite energy must satisfy $w^2\le 1$.
 The precise finite energy restrictions on the behavior of $w$ come
from the nonderivative term in \ymhair, and since $\wwp =0$ gives the relation
$$
w w^{\prime\prime}= {R^2\over r^2}\left(1+w\right)w\left[
w\left(w-1\right)+\mu^2r^2\right]\ \ \ \left(\wwp =0\right),\eqn\ymtp$$
this term also governs the oscillatory properties of finite energy solutions.
{}From \ymtp, we see that $w^{\prime\prime}>0$ for $r>1/2\mu$, so
finite energy solutions must have their final turning points before $r=1/2\mu$
and satisfy $w^{\prime}<0$ after $r=1/2\mu$. For $r<1/2\mu$, on the other hand,
 eqn.\ymtp~
indicates that $w w^{\prime\prime}<0$ in the range
$\left( 1-\sqrt{1-\left( 2\mu r\right)^2}\right) /2
 <w<\left(1+\sqrt{1-\left( 2\mu r\right)^2}\right) /2$, so solutions can
exhibit nontrivial behavior for $w>0$ and still satisfy the
 boundary conditions at infinity.
If $w>\left(1+\sqrt{1-\left( 2\mu r\right)^2}\right) /2$, however, $w$ will
increase beyond $w^2=1$, so this range is forbidden for finite energy
solutions. Since we have used only the $r\rightarrow\infty$ boundary conditions
 and not mentioned initial conditions on $w$ or the metric in this discussion,
the results
$$\eqalign{
r<{1\over 2\mu} :& \>\>\>\> -1<w<{1\over 2}
\left(1+\sqrt{1-\left( 2\mu r\right)^2}\right) \cr
r>{1\over 2\mu} :& \>\>\>\> w^{\prime}< 0 ,\ \ w^{\prime\prime}>0\cr}\eqn\wlim
$$
describe both regular and black hole solutions.

The full boundary conditions may be determined from eqs.\meqpro --\ymproc . For
regular solutions, finite energy density $T_{tt}$ and regularity of $\tr$
at the origin ($ T^{\prime}\br0 =0$) give
\vfill\supereject
$$\eqalignno{
  2\mr &= {\cal O}\left( r^3\right) &\eqname\mzerexpa\cr
  \ln T\brr &= {\cal O}\left( r^2\right) &\eqname\tzerexpa\cr
  w\brr &= -1+ {\cal O}\left( r^2\right) .&\eqname\wzerexpa\cr}
$$

It is significant that $w=-1$ is the only possible initial value; in
EYM theory the  self-interaction term $\ww1^2 / r^2$ in $m^{\prime}$ provides
the two possibilities $w_{\pm} =\pm 1$ [\bart], but the vector field mass in
our theory  excludes $w_{+}$ both at the origin and as $r\rightarrow \infty$.
The $\mu^2$ terms also  breaks
the discrete symmetry $w\rightarrow -w$ of the EYM equations
which gives rise to degenerate mirror-image solutions in [\bart ] and [\biz].
These differences will play an important role in the
spectrum of solutions.

For black hole solutions, the field equations with $m\brh = r_{h}/2$ give
$ m^{\prime}\brh, w^{\prime}\brh$,  and $\delta^{\prime}\brh$ when $\mu $
is specified. Full use of the metric initial conditions allows us to expand
near the horizon:
$$\eqalignno{
  m\brr &= r_{h}/2 +m^{\prime}\brh\rrh+{\cal O}\left( r-r_{h}\right)^2
&\eqname\mrhexpa\cr
  \dr   &= 0+\delta^{\prime}\brh\rrh +{\cal O}\left( r-r_{h}\right)^2
&\eqname\drhexpa\cr
  w\brr &= w\brh + w^{\prime}\brh\rrh+{\cal O}\left( r-r_{h}\right)^2 ,
&\eqname\wrhexpa\cr}
$$
where $r_{h}$ and $w\brh$ are to be chosen to yield a finite energy solution.

Since the only allowable vacuum value of the connection function for both
regular and black hole solutions is $w=-1$, the behavior of
eqs.\meqpro--\ymproc
{}~as $r\rightarrow\infty$ gives
$$\eqalignno{
  \mr &\sim M -\mu c^2 e^{-2\mu r} &\eqname\minfa\cr
  \ln T\brr &\sim \ln\left( {1\over T_{o}}\right) +{M\over r}
 &\eqname\tinfa\cr
  \dr   &\sim -\delta_{o} + \mu c^2
{e^{-2\mu r}\over r} ,&\eqname\dinfa\cr
  w\brr &\sim -1+ c e^{-\mu r}, &\eqname\winfa\cr}
$$
where $c$ is some positive constant and $T_{o}$ and $\delta_{o}$ refer
to the unrescaled initial values introduced in eqs.\tdelb.
 Thus the presence of the $\mu^2 $ terms in ENAP theory give exponentially,
 rather than polynomially, decaying fields as we expect.

{\parindent=0pt\bf  }
{\parindent=0pt\bf Numerical Regular Solutions}

We use a standard one-parameter ``shooting'' method to find regular solutions
to the ENAP equations. The formal power series describing the boundary
conditions \mzerexpa -\wzerexpa ~at $r=0$ is
$$\eqalignno{
  2\mr &= 4b^2r^3 +{2\over 5}\left(-8b^3+3b^2\mu^2\right) r^5 +{\cal O}
\left( r^7\right) &\eqname\mzerexpb\cr
  \ln T\brr &= -2b^2r^2 -{1\over 5}\left( 12b^4-4b^3+\mu^2 b^2\right) r^4 +
{\cal O}\left( r^6\right) &\eqname\tzerexpb\cr
  w\brr &= -1+ br^2 +{1\over 10}\left( 8b^3-3b^2+\mu^2 b\right) r^4 +{\cal O}
\left( r^6\right) ,&\eqname\wzerexpa\cr}
$$
which depend only on $b>0$ and $\mu $ and conform to the EYM results [\bart]
for $\mu =0$. We use these conditions evaluated
at $r=10^{-3}$ as initial data in a standard ordinary differential equation
solver with a global error tolerance of $10^{-12}$, adjusting $b$ for fixed
$\mu $ and integrating outward in an attempt to meet the boundary conditions
 at infinity. The bracketing condition for finite energy solutions is
the same as in [\bart ]: for a small range of $b$ in the vicinity of a
solution,
$w$ approaches its asymptotic value $w\rightarrow -1$ and either crosses
through $w=-1$ and rapidly goes to $w=-\infty$, or experiences a turning point
and rapidly goes to $w=+\infty$. Though $2m\left( r\right)\rightarrow r$
as $\left| w\right| $ diverges, at a discrete value of $b$ in this range
lies a finite energy regular solution with the correct asymptotics. The
existence of such solutions has been proven rigorously only in the $\mu =0$
case
[\yau], but the qualitatively similar numerical behavior for $\mu  >0$
lends strong support for our work.

As in EYM theory, finite energy solutions
are characterized  by the
oscillatory behavior anticipated above in the near-field region $r \gsim 1$.
Distinct solutions are classified by $k$, the number of zeros of $w$,
which increases with increasing $b$ but must be even for $\mu >0$
to conform to the
boundary conditions \winfa~ and \wzerexpa. A striking difference between our
non-abelian Proca solutions and the EYM solutions of [\bart] is that for
$\mu >0$, we find
two distinct classes of even-$k$ solutions.
We will now discuss each class in turn.

One class possesses physical characteristics very similar to the even-$k$
solutions of [\bart] for small $\mu $, reducing to them in the limit
$\mu\rightarrow 0$;
 the presence of the vector field
mass only slightly perturbs the gross behavior of the field.
 Because it
offers the best hope for dynamical stability, we focus our attention on the
$k=2$  solutions. In the range
 $0<\mu<4.454\times 10^{-2}$, the shooting
parameter varies over $0.6517>b>0.5787$  while the width
of the single peak of $w\brr$ increases as the mass increases:
$0.9713< M <0.9936$ (cf. fig.~1a).
For $\mu >4.454\times 10^{-2}$, $w$ does not approach
its asymptotic value before an additional turning point occurs and
$w\rightarrow +\infty$: the $\mu^2$ term in the Proca
equation begins to dominate at a radius $r\sim 1/\mu\approx 20$ before the
 non-abelian self-interaction becomes negligible,
thereby making impossible $k$-node solutions for any value of $b$.
Another way to understand the occurrence of a maximum $\mu $ value
is through eqn.\meqpro;
if  the $\mu^2$ term dominates the energy density,
 the radius will always approach the
Schwarzschild radius $r=2m$ beyond some value of $\mu $
and solutions with smooth geometry will be
forbidden.

The second class of solutions resembles the $(k-1)$ odd-node solutions
of [\bart] as  $\mu\rightarrow 0$, with $w$ approaching the
forbidden asymptotic value $w=+1$ until large $r<1/2\mu$,
where a turning point and then
and additional node occur before $w\rightarrow -1$ (fig.~1b).
 As the mass parameter
increases toward $\mu=4.454\times 10^{-2}$, this type of solution begins to
 resemble the limiting case of the first solution type, until
eventually the two classes converge at the maximum value of $\mu$:
for  $0<\mu<4.454\times 10^{-2}$, the shooting
 parameter and total mass vary over $0.4537<b<0.5787$ and
$0.8286< M <0.9936$, respectively. Thus solutions to ENAP theory
bifurcate into two branches, with the bifurcation point corresponding to
the $\mu$ at which $k$-noded solutions no longer occur.
In shooting parameter space, the bifurcation condition corresponds to the
shrinking of the interval over which $k$ nodes occurs from the maximal value
of [\bart] ($0.4537<b<0.6517$ for $k=2$) to zero, with finite-energy solutions
occurring at the endpoints.

The existence of a second class of solutions and the bifurcation
phenomenon can be understood from a heuristic argument of scales.
In EYM theory, there is one dimensionful parameter $g$ whose units are
$[T][M]^{-1/2}[L]^{-3/2}$ (where for the remainder of this paragraph we
do not set $G = 1$)
which sets the scale of the solutions to be on the order of $G^{1/2} g^{-1}$.
By scale here we refer
to the approximate value of the radius beyond which the fields in the theory
exhibit no nontrivial behavior and rapidly approach their asymptotic values.
In ENAP theory, on the other hand, we have two dimensionful parameters
$[g]$ and $[\mu]$, the latter of whose units is $[L]^{-1}$. There are thus
two distinct length scales : $ G^{1/2} g^{-1}$ and $\mu^{-1}$.
The former sets the
scale for one class of our solutions and the latter sets the scale for our
second class.
Notice that the first scale is dependent on gravitational interactions while
the second is not.
As $1/\mu\rightarrow \infty$, the last node of a $k$-node
 solution
in the second family is pushed off to infinity and the solution approaches
an odd-node solution of the EYM system with its scale being set by the
only remaining parameter, $G^{1/2}g^{-1}$.  These quasi-odd-$k$ solutions
\REF\footb{We shall use the prefix ``quasi-'' to describe solutions
when their  $\mu\rightarrow 0$  behavior suggests a correspondence
to some other solution category.} [\footb ]  therefore have the scale of
their inner structure set by gravity while  the scale of
their asymptotic structure is
set by the Proca mass
\REF\esky{
P.~Bizon and T.Chmaj, ``Gravitating Skyrmions'',
Wien U. preprint UWThPh-1992-23 (1992).}
\REF\footc{
We recently became aware that
a similar bifurcation of solutions has been observed
in the Einstein-Skyrme (ES) system [\esky].} [\footc].

{\parindent=0pt\bf  }
{\parindent=0pt\bf Numerical Black Hole Solutions}

To find numerical black hole solutions, we followed a similar shooting
procedure. The field equations give
$$\eqalignno{
  m^{\prime}\brh &= {\wwrh1^{2}/ 2r_{h}^2} +\mu^2\wrh1^2 &\eqname\mrhb\cr
  w^{\prime}\brh &= {\mu^2\wrh1 r_{h}^2 -\wwrh1 w\brh \over
  r_{h}-{\wwrh1^2 / r_{h}} -2\mu^2\wrh1^2 r_{h}} &\eqname\wrhb\cr
  \delta^{\prime}\brh &= -{2\left[ w^{\prime}\brh\right] ^2 / r_{h}}
&\eqname\drhb\cr}
$$
on the horizon, and we can construct initial data from eqs.\mrhexpa --\drhexpa
 ~with only the
shooting parameter $w\brh$ and the horizon radius $r_{h}$ unspecified.
Following [\biz ], we examine $r_{h}=1$ and chose $\rrh<10^{-12}$ so that the
errors in the initial data are smaller than the global tolerance.

As in the smooth case, we again find two distinct classes of solutions
whose scales are  respectively set by the two dimensionful parameters in
the theory. In appropriate limits, these solutions approach those of
[\biz] modulo one interesting special case to be mentioned below.
For $\mu >0$
we find that  odd-node solutions are possible
in addition to even-$k$ solutions, since the horizon shields  the
singularity which would occur at $w\left( 0\right)=+1$ for such regular
solutions. These $k$-node solutions are classifiable by their
behavior as $\mu\rightarrow 0$; one class reduces to the $k$-node solutions
of [\biz ], while the other approaches the $(k-1)$-node solutions of
[\biz] as the position of the $k$th node moves out to $r=\infty$. The two
solution branches are again joined at a bifurcation point for some
maximum value of $\mu$ for each $k$. The $k=1$ quasi-even-node
case is special because there
exists no corresponding non-abelian solution with $k-1=0$ nodes in [\biz];
instead, this  solution approaches $w\equiv +1$ as $\mu\rightarrow 0$,
which  corresponds to the ordinary Schwarzschild solution.
The (regular and black hole)
analogue of this solution in Einstein--Yang--Mills--Higgs theory
(to be discussed shortly) is significant as it allows us to make contact
with the known existence of a flat space sphaleron.
This is the only set of solutions (black hole {\it or} regular) for which
the limiting case is essentially weak-gravity (weak in the  present
sense meaning that it
becomes arbitrarily close to Schwarzschild).
Hence, ENAP black holes
are a set of fundamentally non-abelian solutions which, like the solutions of
[\biz], possess metrics which interpolate between the RN and
Schwarzschild metrics, but which
include as limiting cases both quasi-Schwarzschild and quasi-RN
($k\rightarrow\infty$ here and in [\biz]) solutions.
In this sense, the spectrum of solutions itself interpolates between
the Schwarzschild and pure-magnetic RN black holes.

Though the solutions for all node numbers exhibit bifurcation, we
focus our attention on the lowest odd-- and even-node solutions.
In figs.~2a--b, the mass and connections for the two branches
of the $k=1$ solutions are plotted. The limiting value of $\mu=0.1233$
joins the solution branches, which are described by $1>w\brh>0.8500$,~
$0.5000<M<1.0052$,~$0<\delta_{o}<0.2854$ (quasi--$k=0$) and
$0.6322<w\brh<0.8500$, ~$0.9372<M<1.0052$,~$0.5485>\delta_{o}>0.2854$.
For $k=2$, the maximum value $\mu=1.752\times 10^{-2}$ gives the
branches $-0.6322<w\brh<-0.5027$,~$0.9372<M<1.0052$,~$0.5485<\delta_{o}<0.5831$
 (quasi--$k=1$) and
$-0.3425>w\brh>-0.5027$,~$0.9938<M<1.0052$,~$0.5932>\delta_{o}>0.5831$
 (cf. figs.~2c--d). For small
$\mu$, the deviation from the Schwarzschild mass $M=0.5$ appears to scale with
$\mu^2$ for the quasi--$k=0$ solutions; this can be understood from \mrhb~,
which for $w\lsim +1$ gives $ m^{\prime}\brh \approx 4\mu^2$ independent of
 $r_{h}$.

As we vary $r_{h}\rightarrow 0$, we find that the even-$k$ black hole
solutions reduce to the regular solutions for the same value of $\mu $ in the
manner of [\biz], with the shooting parameter approaching $w\brh=-1+b\
r_{h}^2$,
where $b$ is the regular solution shooting parameter.
 The value of $\mu$ at which the two solution classes
bifurcate also increases as $r_{h}\rightarrow 0$ from its value at
$r_{h}=1$ to the regular solution bifurcation value; for $k=2$,
$1.752\times 10^{-2}<\mu_{bif}<4.454\times 10^{-2}$ as $0<r_{h}<1$.
The reduction to the regular solutions is sensible when we examine
eqs.\mrhb --\drhb ; despite the presence of the horizon,
$w^{\prime}\brh$ and $\delta^{\prime}\brh$ approach the
leading behavior of the $r\rightarrow 0$ regular solution expansions
with $r$ replaced by $r_{h}$, while the  non-horizon mass contribution
$(2m\brr -r_{h})/(r-r_{h}) = 4\ b^2 r_{h}^2$  mimics the regular
solution  mass-radius ratio $2m\brr /r$ in the same limit.

Although for $\mu\not=0$ there are no regular solution limits which
the odd-$k$ black
holes can approach as $r_{h}\rightarrow 0$, there is limiting
behavior as the horizon continues to shield the $w\br0=+1$ singularity.
The shooting parameter for such solutions behaves as $w\brh=+1-b_{\rm eff}\
r_{h}^2$, where
$b_{\rm eff}$ is now an effective regular shooting parameter, and we can
 determine bifurcation $\mu $ values as well as mass
limits for the solution branches. In the $k=1$ case with $r_{h}=10^{-3}$, for
example,  we find for
$0<\mu <0.2010$ the branches
$0<b_{\rm eff}<0.1329$,~$0<M<0.9289$  (quasi--$k=0$) and
$0.4537> b_{\rm eff}>0.1329$,~$0.8286<M<0.9289$; the latter branch
reduces to the $k=1$ solution of [\bart] while former approaches the
Schwarzschild solution with $M$ decreasing proportional to $r_{h}$.
Because the curvature at the horizon
$R_{\mu}{}^{\mu}=-8\pi T_{\mu}{}^{\mu}=\mu^2(1+w)^2/r_{h}^2$
diverges as $r_{h}\rightarrow 0$, the maximum value of $\mu $ will
presumably  decrease
until it falls off faster than $\sim r_{h}$, which will  bring the
expansions \wrhb -\drhb~ and the non-horizon mass contribution into
the regular solution form (we have not investigated this in detail).
Thus the only legitimate quasi-regular
limiting black hole solutions for odd $k$ occur for $\mu\rightarrow 0$,
but these are significant:
the $r_{h}\rightarrow 0$
odd-node black holes help ``complete'' the  regular spectrum,
in that the regular and quasi-regular
solutions to ENAP theory interpolate between
the Schwarzschild vacuum and the gravitating Dirac monopole [\ortiz ].
As mentioned above, we will find that the
quasi-$k=0$ branch of solutions, which are trivial apart
from a rapid transition
between the minima of the non-abelian self-coupling
$w=+1$ and $w=-1$,
exist with or without  event horizons in EYMH theory and
 correspond to gravitating $SU(2)$ sphalerons.

It should be noted that the divergent behavior of $w$ as $r\rightarrow\infty$
used to shoot  both regular and black hole solutions is characteristic of
integration toward a singular point.
We would observe the same qualitative behavior as $r\rightarrow 0$ or $r_{h}$
 if we had  chosen to integrate
inward from some large value of $r$ using \minfa--\dinfa ~as initial data
and $\{ c,M \}$ as shooting parameters. To help verify the existence of our
solutions, we integrated from both small $r$ and large $r$ to a point
inbetween, where we attempted to match the values of $w,\ w^{\prime}$, and
$m$ by adjusting the shooting parameters $\{b,c,M\}$. Using the
value of $b$ found from our original method, we were able to adjust $\{c,M\}$
to give agreement at the common point to an accuracy comparable to the
global tolerance. We repeated this procedure for EYMH solutions
after including additional  parameters associated
with the Higgs field.

{\parindent=0pt\bf\chapter{ EYMH Theory: Regular and Black Hole Solutions
{}~\hfil~~}}
\let\chapterlabel=4

{\parindent=0pt\bf The Higgs Ansatz}

The starting point for our theory is the addition in eq.$(2.8)$~of a complex
scalar Lagrangian to the ordinary YM Lagrangian. Following the standard model,
 we take the Higgs field to be a complex doublet
$\Phi = (\phi^{+},\phi^{-} )$  with the double-well self-interaction
$$
\VP =\lambda \left( \Phi^{\dagger}\Phi\right)^{2}
-\mu^2\left( \Phi^{\dagger}\Phi\right)+{\rm constant} .\eqn\vorig
$$
The most general complex doublet Higgs may be parameterized
$$
{\Phi}\bbx \equiv {1\over \sqrt{2}}
\exp\left[ -{{\bf\tau}\cdot{\bf\xi}\bbx }\right]
\left(\matrix{ 0 \cr
  h\bbx /r \cr}\right) ,\eqn\phidefc
$$
with $h$ and $\xi^{a}$ arbitrary. A parameterization of the Higgs
field which proves useful in finding
 spherically symmetric EYMH solutions can be achieved
by rewriting $\Phi $ in the form
$$
{\Phi}\bbx ={1\over \sqrt{2}}
\left(\matrix{ \psi_{2}\bbx+ i\psi_{1}\bbx\cr
  \phi\bbx -i\psi_{3}\bbx \cr}\right) ,\eqn\phidefcc
$$
where we have grouped three of the degrees of freedom as a vector ${\bf \psi}$.
Substituting into the EYMH Lagrangian $(2.8)$~gives
$$\eqalign{\leymh = -{1\over 4\pi}\Bigl[ & {1\over 4}\f2
+ {1\over 8}\left( \phi^2+\left| \psi\right|^2 \right)\a2 \cr
+ & {1\over 2} g^{\mu\nu}\left[ \pd_{\mu}\phi\pd_{\nu}\phi +
\left(\pd_{\mu}\psi\right) \cdot\left(\pd_{\nu}\psi\right) \right] +
V\left( \phi^2+\left|\psi\right|^2\right)\cr
+ & {1\over 2} g^{\mu\nu} A_{\mu}\cdot\left( \psi\times\pd_{\nu}\psi+
\psi\pd_{\nu}\phi-\phi\pd_{\nu}\psi \right)\Bigr] ,\cr}\eqn\lagsphere
$$
where $A$ is our original connection {\it ansatz} $(3.7)$ . A useful
{\it ansatz} for the Higgs field which yields a spherically symmetric energy
density is to take
$\phi =\phi\left( r,t\right)$ and
 ${\bf \psi}=\psi\left( r,t\right) {\wdhat n_r}$,
with ${\wdhat n_r}$ a unit vector in the radial direction.

The static field
equations in the gauge $b\equiv 0$, without the {\it ansatz}
 $a\equiv d\equiv 0$,
 are
$$\eqalignno{
&{d\over dr}\left[ r^2{T\over R} a^{\prime}\right]
-2RT\left[ \left( w^2+d^2\right) +{1\over 8} r^2\left( \phi^2+
\psi^2\right)\right] a =0 &\eqname\faeqn\cr
&{d\over dr}\left[ {w^{\prime}\over RT}\right] +{\left[ 1-\left(w^2+d^2\right)
\right]\over r^2} w {R\over T}
-{1\over 4}\left[ \w1\phi^2-\left(1-w\right)\psi^2\right]{R\over T} +a^2 w RT
=0 &\eqname\fweqn\cr
&{d\over dr}\left[ {d^{\prime}\over RT}\right] +{\left[ 1-\left(w^2+d^2\right)
\right]\over r^2} d {R\over T}
-{1\over 4}\left[ \left(\phi^2+\psi^2\right) d-2\phi\psi\right]{R\over T}
+a^2 d RT
=0 &\eqname\fdeqn\cr
&{d\over dr}\left[{r^2\phi^{\prime}\over RT}\right]
-V^{\prime}_{\phi} r^2 {R\over T}
-{1\over 2}\left(\left[\w1^2 +d^2\right]\phi -2d\psi\right){R\over T}
+{1\over 4}a^2r^2\phi RT =0 &\eqname\fphieqn\cr
&{d\over dr}\left[{r^2\psi^{\prime}\over RT}\right]
-V^{\prime}_{\psi} r^2 {R\over T}
-{1\over 2}\left(\left[\left(1-w\right)^2
 +d^2\right]\psi -2d\phi\right){R\over T}
+{1\over 4}a^2 r^2\psi RT =0 ,&\eqname\fpsieqn\cr
}
$$
with the constraint equation
$$
{2\over RT}\left[\left(d^{\prime}w-w^{\prime}d\right) +{1\over 2}r^2\left(
\phi^{\prime}\psi-\psi^{\prime}\phi\right)\right] =0 \eqn\bconstr
$$
arising from the gauge choice.
Following Witten [\wit], it is useful to express the $w$ and $d$
degrees of freedom
in  complex scalar form:
$$
w\brr -id\brr = f\brr \exp i\alpha\brr .\eqn\wdscalar
$$
Substitution of $w=f\cos\alpha$ and $d=-f\sin\alpha$
into the constraint equation with $\phi\equiv\psi\equiv 0$, $v\equiv 0$
gives the result $f^2\alpha^{\prime}=0$, implying that the  gauge
choice allows for the dynamical elimination of one of the remaining degrees
of freedom: $w$ and $d$ are related by a multiplicative constant (the
equations for $d$ and $w$ in EYM theory are identical),
 and $d\equiv 0$ in
[\bart ] and [\biz ] is not
an {\it ansatz} but a result of the  choice  of the constant $\alpha\equiv 0$
[\bart, \biz]. With $\phi$ and $\psi$ nonzero, the constraint equation
becomes
 $$
2f^2\alpha^{\prime}+{1\over 4}h^2\gamma^{\prime}=0.\eqn\bconstrb
$$ Examination of the field equations then implies that if either $\alpha$
or $\gamma$ is chosen to be a constant, then $\alpha - \gamma$ = $n\pi$.

We will only study the case in which $\alpha$ and $\gamma$ are  constants, and
hence
obtaining static solutions to EYMH theory
reduces to  solving the coupled
equations for $a,f,$  and $h/r$ and choosing $\alpha$ and $n$.

The equation for $a$ may be rewritten
$$
{1\over 2}{d\over dr}\left[ r^2{T\over R} \left(a^2\right)^{\prime}\right]=
r^2{T\over R} \left(a^{\prime}\right)^2+
2RT\left[f^2+{1\over 8} h^2\right] a^2 ,\eqn\faeqnb
$$
while the $f$ and $h/r$ equations become
$$\eqalignno{
&{d\over dr}\left[ {f^{\prime}\over RT}\right] +{\left[ 1-f^2\right] f\over
r^2}
{R\over T}
-{1\over 4}\hofr^2\left[ f+\cos n\pi\right]{R\over T}+a^2fRT =0&\eqname\feqn\cr
&{d\over dr}\left[{r^2\left( h /r\right)^{\prime}\over RT}\right]
-V^{\prime}_{h} r^2 {R\over T}
-{1\over 2}\left[f+\cos n\pi\right]^2\hofr {R\over T}
+{1\over 4}a^2r^2\hofr RT =0. &\eqname\heqn\cr
}$$
For odd $n$, the above equations can be put in the same form
as the even-$n$ equations by defining ${\wdtil f}\equiv -f$, so the
solutions for odd and even $n$ are isomorphic despite the fact that the
physical field configurations are quite different. Before exploring these
differences,  we note that the $r\rightarrow\infty$ boundary conditions
on $f$ and $h$ require $a\brinf =0$, and for regular solutions $a\br0 =0$.
Since the right side of \faeqnb~ is positive semi-definite, we must have
$a\equiv 0$ for static regular solutions.
\REF\galers{A.A.~Ershov and D.V.~Gal'tsov\journal Phys.Lett.&A~150(1990)159.}
 Thus
the no-dyon results  for EYM theory [\galers] extend to EYMH theory.

Though solutions to \feqn--\heqn~do not depend explicitly on the values
of $\alpha $ and $n$, some physical properties of the Higgs field and
connection
are affected by these parameters.
Consider $\alpha =0$, for example. The even-$n$ solutions reduce to our
{\it ansatz}, while the odd-$n$ solutions have
have  a ``hedgehog'' Higgs field
$\left(\phi,\psi\right) \equiv\left( 0,\pm h/r\right)$.
We will shortly see that one of the smooth $k=1$ node solutions  of this form
which we find in the next
section has a limit in which it becomes close to the flat space sphaleron
of YMH theory.

{\parindent=0pt\bf }
{\parindent=0pt\bf EYMH Equations and Boundary Conditions}

To obtain static, spherically symmetric solutions we assume
$\phi=\phi\brr=v+\eta\brr$ and $\psi = 0$.
The Lagrangian
for the theory may then be written in the form
$$
\leymh =  -{1\over 4\pi}
\left( {1\over 4}\f2 + {1\over 2}\left({g\phi\over 2}\right)^2\a2
+{1\over 2}\dele+\Vp\right) ,\eqn\eleymhc
$$
where $\Vp =\lambda\left[\phi^2 -v^2\right]^2/4$
 satisfies $V\left(\pm v\right) =0$.

Specializing our results of the last section, the gauge field equation is
$$
{d\over dr}\left( {w^{\prime}\over RT}\right) +{w\ww1\over r^2}{R\over T}
-\left({g\phi\over 2}\right)^2\w1 {R\over T} =0,\eqn\ymha
$$
and the Higgs equation is
$$
{d\over dr}\left({r^2\eta^{\prime}\over RT}\right)-{1\over 2}\w1^2\phi{R\over
T}
-\Vpp r^2 {R\over T} =0 ,\eqn\higgsa
$$
where $V^{\prime}\equiv dV/d\phi=dV/d\eta$.

{}From variation of the full action with respect
to $g_{\mu\nu}$ we obtain the energy-momentum tensor,
$$\eqalign{
8\pi T_{\mu\nu} = 2F_{\mu\gamma}F_{\nu}{}^{\gamma} -&{1\over 2}
g_{\mu\nu}\f2 +\left[
2\left({g\phi\over 2}\right)^2  A_{\mu} A_{\nu}
-g_{\mu\nu}\left({g\phi\over 2}\right)^2 \a2\right] \cr
 +& 2\pd_{\mu}\eta\pd_{\nu}\eta -g_{\mu\nu}\left[ \dele +2\Vp\right] ,\cr}
\eqn\tabeymh
$$
and the $(tt)$ and $(rr)$ Einstein equations:
$$\eqalignno{
m^{\prime} =&\R2\left( {w^{\prime}\over g}\right)^2 +{1\over 2}{\ww1^2\over g^2
 r^2}+\left({\phi\over 2}\right)^2\w1^2 &\eqname\meqeymh\cr
&\phantom{\R2\left( {w^{\prime}\over g}\right)^2}\;
+{1\over 2}\R2\left( r\eta^{\prime}\right)^2 +\Vp r^2 &\cr
r\R2{T^{\prime}\over T} = &-\R2\left( {w^{\prime}\over g}\right)^2 +
{1\over 2}{\ww1^2\over g^2r^2} +\left({\phi\over 2}\right)^2\w1^2
&\cr
&-{1\over 2}\R2\left( r\eta^{\prime}\right)^2 +\Vp r^2 -{m\over r}.
&\eqname\teqeymh\cr}
$$
For black hole solutions, we again replace the auxiliary
$T^{\prime}$ equation with an equation for $\delta^{\prime}$,
$$
\delta^{\prime} =-\left[ 2\left( w^{\prime}/ g\right)^2 +
\left( r\eta^{\prime}\right)^2\right] /r .\eqn\deqeymh
$$
The Einstein equations can once again be used to express the equations
of motion in a form independent of $g_{tt}$:
$$\eqalignno{
&r^2\R2 w^{\prime\prime}+
\left[ 2m-{\ww1^2\over g^2 r}-2\left( {\phi\over 2}\right)^2\w1^2r
-2\Vp r^3\right]w^{\prime} &\cr
&\phantom{r^2\R2w^{\prime\prime}}\;
+\ww1 w -\left( {g\phi\over 2}\right)^2\w1 r^2 =0 &\eqname\ymhiggs\cr
&r^2\R2\eta^{\prime\prime}+
\left[ 2\left( r-m\right)-{\ww1^2\over g^2r}
-2\left( {\phi\over 2}\right)^2\w1^2r
-2\Vp r^3\right]\eta^{\prime} &\cr
&\phantom{r^2\R2 \eta^{\prime\prime}}\;
-{1\over 2} \w1^2\phi  -\Vpp r^{2} =0. &\eqname\higgsb\cr}
$$

The addition of the Higgs field and new coupling constants $\lambda ,\mu^2$
requires a reexamination of the scaling properties of the field equations.
{}From the replacement of $\mu^2$ in the ENAP theory with
$\left(g\phi /2\right)^2$  and examination of $\leymh $, we find
$$
\left[ \eta\right]=\left[ v\right]= 1, \>\>\>\>\>
\left[ \lambda\right]^{1/2} =\left[ \mu\right]=\left[ L\right]^{-1},
\eqn\dimanalb
$$
so by introducing ${\wdhat \lambda}=\lambda /g^2$ and
${\wdhat \mu}=\mu /g $, we obtain the same overall scaling behavior of the
Lagrangian found in section III:
$$
\leymh = g^2\leymh\left(\rhat ,\lambhat,\muhat \right) .\eqn\leymhscale
$$
Solutions for $g\not= 1$ will again be related to dimensionless quantities as
before,
with the addition of $\eta_{g} =\eta\bgrr $, and the dimensionless
field equations can be obtained by replacing
$\left( r ,m ,\lambda,\mu \right)$  with
$\left(\rhat ,\mhat,\lambhat,\muhat \right)$  and setting $g=1$
in eqs.\meqeymh --\higgsb . We take $g=1$ for the remainder of the paper
and consider all quantities dimensionless unless otherwise specified.

By following the analysis of section III, we can again predict general
features of the solutions and boundary conditions. From the expression for
$\delta $ \deqeymh~, we see that $R/T$ again increases  (and $\delta$ for black
holes {\it decreases}) monotonically with
radius, but because $T$ now satisfies
$$
{d\over dr}\left[{r^2\over R}\left( {1\over T}\right)^{\prime}\right]
= 2\R2\wp2 {R\over T} +\left( {R\over r^2 T}\right)\left[
\ww1^2-2\Vp\right] ,\eqn\tconeymh
$$
it is no longer clear whether $T$ decreases monotonically for regular
solutions. The condition $T>R\ge 1$, with $R$ possessing at least one maximum,
still holds for regular solutions.

The YM equation may be rewritten
$$
{1\over 2}{d\over dr}\left[ {\left( w^2\right)^{\prime}\over RT}\right] =
{\wp2\over RT}+{R\over r^2 T}\left[\left(w^2-1\right) w^2+
\left({\phi\over 2}\right)
^2\w1 wr^2\right] ,
\eqn\ymhhair$$
in general, while at turning points we have the relation
$$
w w^{\prime\prime}= {R^2\over r^2}\left(1+w\right)w\left[
w\left(w-1\right)+\left({\phi\over 2}\right)^2r^2\right]
\ \ \ \left(\wwp =0\right).\eqn\ymtp$$
Since $\phi\brinf\not= 0$
for a spontaneously broken theory,  $w=-1$ is again the
only acceptable asymptotic value  and $w^2<1$ is required for finite
energy solutions. Because $\phi $ is a field and not a constant,
determining restrictions on the value of $w$ and the occurrence of
turning points is not as simple as for ENAP theory. If we assume that
$\phi$ is ${\cal O}\left( v\right)$ in the region $r\gsim 1$, we can again use
a characteristic radius to define two regimes of interest and apply
the analysis of section III:
$$\eqalign{
r\lsim {\cal O}\left( 1/v\right) :& \>\>\>\> -1<w<{1\over 2}
\left(1+\sqrt{1-\left( v r\right)^2}\right) \cr
r\gsim {\cal O}\left( 1/v\right) :& \>\>\>\>\>\> w^{\prime}< 0 ,
\ \ w^{\prime\prime}>0\cr}\eqn\wlimymh
$$
quantify the finite-energy constraints on both regular and black
hole solutions.

To better understand the expected behavior of $\phi $, we rewrite the Higgs
equation
$$
{1\over 2}{d\over dr}\left[  { r^2 \left(\phi^2\right)^{\prime}\over RT}\right]
={\left( r\phi^{\prime}\right)^2\over RT}+{R\over  T}r^2\left[
{\w1^2\over 2r^2}\phi^2 +\Vpp\phi\right] .
\eqn\hhair$$
The obvious requirement for finite energy solutions is $\Vpp\phi \le 0$, which
implies that $\phi $ is restricted to lie between the minima of the potential:
$\left| \phi\right|\le v$. We also note that $\phi =\pm v, 0$ are the only
allowed values as $r\rightarrow \infty $, but because $\Vp$ is nonzero
at $\phi\left(\infty\right) =0$, finite energy  solutions must have
$\phi\left( \infty\right) =\pm v$.   From \hhair,
the equation governing the oscillatory properties of $\phi $ is
$$
\phi \phi^{\prime\prime}= {R^2}
\left[{\w1^2\over 2r^2}\phi^2 +\Vpp\phi\right]
\ \ \ \left(\phi^{\prime} =0\right) .\eqn\htp$$
When the gauge field coupling term is negligible, we see that
$\phi\phi^{\prime\prime} < 0$, which is characteristic of oscillations about
$\phi =0$ (a solution of infinite energy) unless the initial value of
$\phi^{\prime}$ in \hhair~ is large enough for the field to reach one of the
minima of $\Vp$ as $\phi^{\prime}\rightarrow 0$. For regular solutions, we
will find that $\phi^{\prime}\br0 =0$ for $\phi\br0\not= 0$, so
oscillatory behavior will occur unless the gauge field coupling becomes
important. For black hole solutions and regular solutions with $\phi\br0 =0$,
the initial derivative of $\phi $ is nonzero, but the gauge field
term again plays an important role in avoiding  infinite energy solutions.
The behavior of the scalar field is consistent with the analogue of a
particle moving in a potential proportional to $-V $, with the gauge
field interaction and gravitational forces fighting the restoring force of the
potential. These heuristic considerations are in agreement with the
no-hair proof for the Goldstone theory [\adl ], which approaches the
existence of finite energy solutions rigorously but for which the analysis
fails in the presence of a non-abelian gauge field. Thus we expect the
Higgs field  either to smoothly transit  the potential until
$\left| \phi\right| = v$, or to move initially toward $\phi =0$ and
 then turn and roll under the influence of the gauge field toward
$\left| \phi\right| = v$. The $\phi\rightarrow -\phi$
symmetry of the field equations  also implies that every solution
will have a mirror-image solution; we focus here on solutions with
$\phi\brinf =+v$ without loss of generality.

The presence of the Higgs field in EYMH theory allows for two possible
sets of initial conditions for regular solutions. Finite $T_{tt}$  and
regularity of the metric at $r=0$ give
$$\eqalignno{
  2\mr &= {\cal O}\left( r^3\right) &\eqname\mzerexpc\cr
  \ln T\brr &= {\cal O}\left( r^2\right) &\eqname\tzerexpc\cr
  \left(\matrix{
  w\brr \cr
  \eta\brr \cr}\right)
&=\left(\matrix{
 -1+ {\cal O}\left( r^2\right) \cr
 \eta_{o}+ {\cal O}\left( r^2\right)\cr}\right)\ \ \
 {\rm Even-k} &\eqname\wetazerexpa\cr
\left(\matrix{
  w\brr \cr
  \eta\brr \cr}\right)
&=\left(\matrix{
 +1+ {\cal O}\left( r^2\right) \cr
 -v+ {\cal O}\left( r\right)\cr }\right)\ \ \
 {\rm Odd-k}, &\eqname\wetazerexpb\cr
}
$$
where  $k$ again denotes the number  of $w$ nodes and
$-v < \eta_{o} < 0$ for $\phi\brinf =+v$.
Black hole solutions again possess expansions like eqs.$(3.29)$ --
$(3.31)$~ near the horizon, with  the addition of
$$
  \eta\brr= \eta\brh + \eta^{\prime}\brh\rrh+{\cal O}\left( r-r_{h}\right)^2 ,
\eqn\etrhexpa
$$
where $\eta\brh$  brings to three the number of unknowns.

The vacuum values $w\brinf =-1$ and $\eta\brinf =0 $ are shared by
black hole and regular solutions; the behavior of the field equations as
$r\rightarrow \infty$ gives

$$\eqalignno{
  \mr &\sim M -{1\over 2}a^2\left(\sqrt{2}\mu r
\right) re^{-2\sqrt{2}\mu r} &\eqname\minfb\cr
  \ln T\brr &\sim \ln\left( {1\over T_{o}}\right) +{M\over r}
 &\eqname\tinfb\cr
  \dr   &\sim -\delta_{o} + {1\over 2} a^2\left( \sqrt{2}\mu r \right)
e^{-2\sqrt{2}\mu r} &\eqname\dinfb\cr
  w\brr &\sim -1+ c e^{-(v/2) r} &\eqname\winfb\cr
  \eta\brr &\sim  a e^{-\sqrt{2}\mu r} ,&\eqname\etainfb\cr}
$$
where $c$ and $a$ are constants, $v/2$ is the gauge field mass after symmetry
 breaking and $\sqrt{2}\mu$ is the Higgs field mass.

{\parindent=0pt\bf  }
{\parindent=0pt\bf Numerical Regular Solutions}

The presence of the Higgs field makes the solution of the EYMH equations
a two-parameter shooting problem. For even-$k$ solutions, the boundary
conditions \mzerexpc-\wetazerexpa~ become
$$\eqalignno{
  2\mr &= \left(4b^2 +{2\over 3}V_{o}\right)r^3 +&\eqname\mzerexpd\cr
&\phantom{=}\;
{2\over 5}
\left[-8b^3+\left( 3\left({\phi_{o}\over 2}\right)^2 +{16\over 3} V_{o}
\right) b^2+8\left( {1\over 6}V^{\prime}_{o}\right)^2\right]r^5 +{\cal O}
\left( r^7\right) &\cr
  \ln T\brr &= -\left(2b^2 -{1\over 3} V_{o}\right)r^2 &\eqname\tzerexpd\cr
 &\phantom{=}\;
-\left({1\over 5}\left[ 12b^4-4b^3+\left( \left({\phi_{o}\over 2}\right)^2
+4V_{o}\right)b^2 -3\left( {1\over 6}V^{\prime}_{o}\right)^2\right]
 -{1\over 9}V_{o}^2\right) r^4 +
{\cal O}\left( r^6\right) &\cr
  w\brr &= -1+ br^2 +{1\over 10}\left[ 8b^3-3b^2+\left(\left( {\phi_{o}\over 2}
\right)^2+ 4V_{o}\right) b\right] r^4 +{\cal O}
\left( r^6\right) &\eqname\wzerexpd\cr
  \eta\brr &= \eta_{o}+ {1\over 6}V_{o}^{\prime}r^2 +
\left[{1\over 120}V_{o}^{\prime}\left( 24b^2+{20\over 3} V_{o}+
V_{o}^{\prime\prime}\right) +{1\over 40} \phi_{o} b^2\right]r^4 +{\cal O}
\left( r^6\right),  &\eqname\etazerexpd\cr
}
$$
where $V_{o}$, $V_{o}^{\prime}$ and
$V_{o}^{\prime\prime}$ are the potential and its derivatives with respect to
 $\phi$ at  $\phi=\phi_{o}\equiv v+\eta_{o}$, and the shooting parameters are
$b>0$ and $\eta_{o} $. For odd-$k$ solutions, the expansions are
$$\eqalignno{
  2\mr &= \left(4b^2 +{2\over 3}V_{o} +\etanp^2\right)r^3 +&\eqname\mzerexpe\cr
&\phantom{ =}\;{2\over 5}
\left(-8b^3+{16\over 3} V_{o}b^2+
\etanp^2\left[6b^2-3b+\left(\etanp^2+V_{o}+V_{o}^{\prime\prime}\right)\right]
\right)r^5 +{\cal O}
\left( r^7\right) &\cr
  \ln T\brr &= -\left(2b^2 -{1\over 3} V_{o}\right)r^2 +{1\over 9}
 V_{o}^2 r^4 &\eqname\tzerexpe\cr
            &\phantom{ =}\;
-{1\over 5}\left( 12b^4-4b^3+4V_{o}b^2 +{1\over 4}\etanp^2\left[
24b^2-2b-{8\over 3}V_{o}-V_{o}^{\prime\prime}\right]\right) r^4 +
{\cal O}\left( r^6\right) &\cr
  w\brr &= +1- br^2 -{1\over 10}\left(8b^3-3b^2+
 4V_{o}b-{1\over 2}\etanp^2\left[1-8b\right]\right) r^4 +{\cal O}
\left( r^6\right) &\eqname\wzerexpe\cr
  \eta\brr &=  \eta_{o}^{\prime} r  +
{1\over 10}\eta_{o}^{\prime}\left( 8b^2-2b+3\etanp^2+
+{8\over 3} V_{o}+V_{o}^{\prime\prime}\right) r^3 +{\cal O}
\left( r^5\right), &\eqname\etazerexpe\cr}
$$
where $\eta_{o}^{\prime}$ replaces $\eta_{o}$ as the second shooting parameter.
We again evaluate the initial conditions at $r=10^{-3}$ and use global error
tolerance $10^{-12}$ in a standard ordinary differential
equation solver, adjusting  either $\left(b,\eta_{o}\right)$  or
$\left(b,\eta_{o}^{\prime}\right)$ for fixed $\lambda$ and $\mu$
 and integrating toward $r=\infty $. The finite energy solution
bracketing condition for $b$
is similar to that of ENAP theory: for $\eta_{o}$ or $\eta_{o}^{\prime}$ in the
vicinity of a solution, there exists a range of $b$
for which $w\rightarrow -1$ but then rapidly  approaches
$\left|w\right| =\infty $, with a discrete value of $b$ giving the correct
asymptotics for $w$. As we adjust $\eta_{o}$ or $\eta_{o}^{\prime}$, the
Higgs field either passes through $\eta=0$ and diverges or experiences a
turning point and begins oscillating about $\eta=-v$, with
discrete values of $\left( b,\eta_{o}\right)$
or $\left( b,\eta_{o}^{\prime}\right)$ giving the monotonic approach
of both fields toward the boundary values as $r\rightarrow \infty$.
 The difficulty of the two-dimensional shooting problem was
compounded by the presence of the two free parameters $\lambda $ and $\mu$.
Examination of \winfb--\etainfb reveals that
a significant disparity between $\sqrt{2}\mu $ and
$v /2$ makes the accurate determination of solution shooting parameters
very difficult, since both fields must simultaneously approach their
asymptotic values for both bracketing conditions to occur. As a consequence, we
focus on solutions with $\sqrt{2}\mu =v/2$ (corresponding to $\lambda =1/8 $).

The general properties of finite energy regular solutions are the same as for
 ENAP theory. Solutions are again characterized by $w$ oscillations in the
region $r\gsim 1$ and classified by the node number $k$, which may now be
odd as well as even. For odd $k$, the Higgs deviation monotonically increases
in the range $-v\le\eta\le 0$, while for even $k$ it undergoes a turning
point very close to $\eta_{o}$ and then monotonically increases to $\eta\brinf
=0$. Two distinct sheets of solutions for each $k$ again arise
(for any choice of $\lambda$ or $\mu$ in the appropriate range
we have two solutions for each choice of $k$), but for
$\lambda = 1/8$ they do not precisely converge, and  a different maximum
value $v_{\rm max}$ for each sheet may be determined.
Though
we have not fully investigated this phenomenon, these two
sheets corresponding to the different solution  classes
presumably bifurcate at a point with $\lambda_{bif}\not= 1/8$.
The results for $k=1$ and $k=2$ are shown in fig.~$3a-d$. As $v$ varies in the
range $0<v<0.599$ for the first $k=1$ branch, the shooting parameters vary
over $0.454>b>0.261$ and $0.323v<\eta_{o}^{\prime}<0.329 v_{\rm max}$ and
the mass behaves as $0.8286<M<0.9272$;  the $v\rightarrow 0$ limit is the
$k=1$ solution of [\bart]. The quasi-$k=0$ branch is described by
 $0<v<0.619$,
$0.292v^2 < b< 0.206$, $0.434v^2<\eta_{o}^{\prime}<0.307 v_{\rm max}$, and
 $1.821v<M<0.9380$. This  approaches the weak-gravity regime
quickly as $v$ decreases: $g_{tt}\br0$ increases from its minimum value
$0.1557$ at $v_{\rm max}$ to
$g_{tt}\br0 =0.9998$ at $v=10^{-2}$. We also note that the scaling of the
mass with $v$ is characteristic of sphaleron solutions to YMH theory, but
address the relationship between our solutions and sphalerons more fully below.
For $k=2$, the two solution branches are described by
$0<v<0.120$,\ $0.652>b>0.585$, \
$-.870v<\eta_{o} <-.845 v_{\rm max}$, \ $0.9713<M<0.9972$
and
$0<v<0.122$,\ $0.454<b<0.568$, \
$-v<\eta_{o} <-.850 v_{\rm max}$, \ $0.8286<M<0.9982$ (quasi-$k=1$).
As $v\rightarrow 0$, the branches approach the $k=2$ and $k=1$ solutions
of [\bart], with the initial Higgs deviation of the quasi-$k=1$ branch
approaching the odd-$k$ initial value $\eta_{o}=-v$.

We can again understand in heuristic terms the bifurcation of the solution
space into two branches by appealing to length scales
\REF\footd{We note that in
EYMH theory, we have only examined the one-- and two-node solutions and hence
it is possible that our discussion is sensitive to special features
of these limited cases.}
[\footd ] .
 The coupling
parameters in this model provide us with two length scales:
$ L_1 = g^{-1} G^{1/2}$ and
$L_2 = g^{-1} \sqrt{\lambda / \mu^2}$ (where we do not
set G = 1 for the remainder of this paragraph). As in our discussion
of ENAP theory, the former of these scales is set by gravity while the
latter is not.  One branch of our solutions has its scale (the position
after which the fields rapidly approach their asymptotic values) set by
$L_1$ and the latter by $L_2$. In the limit that the $v$ goes to zero
the k node solutions on the latter branch have their last node pushed out
to infinity and they go over to $k - 1$ node solutions with scale set by
$L_1$. An interesting special case is that of our $k = 1$ node solutions
(smooth or black hole) since they would appear to approach a $k = 0$ node
solution. This means that their scale is set totally by $L_2$ and
hence can exist with arbitrarily weak gravity. In particular, our smooth
solution of this sort can be related to the flat space sphaleron in this limit.
Explicitly, taking our solution with the choice $ \alpha = 0 $ and
odd--$n$ we can write the solution in the form
$$\eqalignno{
A_{\mu}&={\ff\over 2}U\partial_{\mu}U^{-1} &\cr
\Phi&={1\over \sqrt{2}}\hofr U\left[\matrix{ 0\cr 1\cr}\right]&\eqname\asph\cr
U&=\exp\left[ -n\pi \tar\right], &\cr}
$$
which is physically equivalent to the YMH sphaleron
{\it ansatz} of
\REF\dhn{R.F.~Dashen, B.~Hasslacher, and A.~Neveu\journal Phys.Rev.&D~10(1974)
4138.}
\REF\klink{F.R.~Klinkhamer and N.S.~Manton\journal Phys.Rev.&D~30(1984)2212.}
[\dhn ] and [\klink]. Though the sphaleron solutions utilizing \asph~
have only been found  for $k=1$ in YMH theory, we see that they appear to
exist for all $k$ when we include gravity.

{\parindent=0pt\bf  }
{\parindent=0pt\bf Numerical Black Hole Solutions}

To find numerical black hole solutions, we used the conditions
$$\eqalignno{
  m^{\prime}\brh &= {\wwrh1^{2}/ 2r_{h}^2} +\left[\phi\brh\over 2\right]^2
\wrh1^2 +V\left(\phi\brh\right) r_{h}^2&\eqname\mrhc\cr
  w^{\prime}\brh &= {\left[\phi\brh /2\right]^2\wrh1 r_{h}^2
-\wwrh1 w\brh \over
  r_{h}-{\wwrh1^2 / r_{h}} -2\left[\phi\brh /2\right]^2
\wrh1^2 r_{h}-2V\left(\phi\brh\right) r_{h}^3 } &\eqname\wrhc\cr
  \eta^{\prime}\brh &= {\wrh1\left[\phi\brh /2\right] +V^{\prime}
\left(\phi\brh\right) r_{h}^2 \over
  r_{h}-{\wwrh1^2 / r_{h}} -2\left[\phi\brh /2\right]^2
\wrh1^2 r_{h}-2V\left(\phi\brh\right) r_{h}^3} &\eqname\etarhc\cr
  \delta^{\prime}\brh &= -\left( 2\left[ w^{\prime}\brh\right]^2 +
\left[ r_{h}\eta^{\prime}\brh\right]^2 \right)/ r_{h}
&\eqname\drhd\cr}
$$
on the horizon, and use $w\brh$ and $\eta\brh$ as shooting parameters for
$r_{h}=1$. The results for $k=1$ and $k=2$ are shown in figs.~$4a-d$.

Two solution branches again appear for each $k$, with the quasi-$k=0$  branch
distinguished by its weak-gravity $v\rightarrow 0$ limit.
The $k=1$ solutions are decribed by $0<v<0.356$,~ $(1-0.292v^2)>w\brh> 0.870$,
{}~ $-(1-0.217v)v<\eta\brh<-0.304$,
{}~ $(0.5+1.820v)<M<1.0181$,~$1.19v^2<\delta_{o}< 0.3310$ (quasi--$k=0$) and
$0<v<0.331$,~ $0.632<w\brh< 0.801$,
{}~ $-0.826v >\eta\brh>-0.275$,
{}~ $0.9372<M<1.0043$,~$ 0.5485>\delta_{o}>0.4300$. The specifications
of the  $k=2$ solutions are
$0<v<0.0486$,~ $-0.632<w\brh< -0.518$,
{}~ $-(1-1.62v)v>\eta\brh>-0.0429$,
{}~ $0.9372<M<1.0073$,
{}~$0.5485<\delta_{o}< 0.5799$ (quasi--$k=1$) and
$0<v<0.0475$,~ $-0.345>w\brh>-0.489$,
{}~ $-0.901v >\eta\brh>-0.0417$,
{}~ $0.9938<M<1.0066$,~$ 0.5932>\delta_{o}>0.5844$.

We see that these black hole solutions have nontrivial gauge and Higgs
field structure outside of the horizon. This could not happen if the gauge
group were abelian -- it relies, as discussed in section II, on the non-abelian
nature of the field theory being studied here.

{\parindent=0pt\bf\chapter{Conclusions~~\hfil~~}}

In this paper we have studied spherically symmetric
classical solutions to $SU(2)$ non-abelian
Proca theory and spontaneously broken gauge theory. Our main intent has been
to exploit a gap in the known no-hair results -- they do not necessarily apply
to nonlinear field theories --  to find black hole solutions,
which have nontrivial
structure outside of the horizon, in the most
familiar and relevant kinds of field theories . In particular, we have
presented strong
numerical evidence indicating that such field theories, when coupled to
Einstein gravity,
  do admit spherically symmetric black hole solutions in which
the gauge and Higgs fields decay to their asymptotic values exponentially
far from the hole, in contrast to previous expectations.

An important physical question is whether these solutions are stable
(although we reemphasize that the no-hair proofs have to do with existence
of classical solutions, not with their stability).
We have not exhaustively studied this question, but it does seem somewhat
unlikely that  they are stable. Our EYMH solutions have a fairly direct
relation to sphaleron solutions (even though many of them involve
strong gravity) and hence are probably unstable. In fact, the arguments of
\REF\galvol{D.V.~Gal'tsov and M.S.~Volkov\journal Phys.Lett.&A~162(1992)144.}
[\straua--\straub] and  [\galvol] as applied to the smooth and black hole
solutions
of EYM solutions  [\bart ,\biz] seem likely ensure that our solutions are
unstable. At this time we have only definitively studied this question for
the smooth lowest node ENAP solutions using linear stability analysis
along the lines of [\straua] which does in fact show these solutions to
be unstable.
It would be of interest to carry out this analysis fully, especially in light
of the natural suggestion [\straub ] that a no--hair theorem might hold
if one demands stability. It is also important to develop methods
to deal with the latter possibility in a general setting.

Another interesting question is to try to understand at a more fundamental
level why these smooth and black hole solutions exist. Recently,
the authors of \REF\wald{D.~Sudarsky and R.~Wald,
``Extrema of Mass, Stationarity and Staticity, and Solutions to the
Einstein-Yang-Mills Equations'', U. of Chicago preprint (1992).}
 [\wald ] have given some very interesting
plausibility arguments to argue for the existence of the solutions of
[\bart] and [\biz]. These arguments rely on properties of the Yang-Mills
configuration space associated with the existence of large gauge
transformations. However, one can evade the no-hair arguments in
as simple a theory as two scalar fields so long as the quartic
potential is chosen judiciously. It would be interesting to see if
smooth and black hole solutions can be found in such models as this
will help to clarify the essential physics. We are presently studying this
question and will report on it elsewhere.

\FIG\fa{Two-node regular solutions to Einstein-Non-Abelian-Proca theory. The
connection $\left( 1+w\brr\right)$ and total mass
are plotted as functions of radius
for a range of vector field mass values $0<\mu<\mu_{\rm max}$.
Increasing $\mu$ in figures (1a) and (1b)  (and in figure (2) )
corresponds to decreasing the
value of the radius at which the solution exponentially decays to its
vacuum value.
 The $k=2$ and
 quasi--$k=1$ solutions, so
named because as $\mu\rightarrow 0$ they resemble the $k=1$ solution of Bartnik
and McKinnon [\bart], bifurcate at $\mu=\mu_{\rm max}$.}

\FIG\fb{
One- and two-node black hole solutions to Einstein-Non-Abelian-Proca
theory for horizon radius $r_{h}=1$. Like the regular case,
both odd and even-node
solutions exhibit bifurcation at a maximum value of $\mu $. The quasi-$k=0$
solutions resemble the Schwarzschild solution $w=1$ as $\mu\rightarrow 0$;
this is the only class of ENAP solutions with a weak-gravity limit.}

\FIG\fc{One- and two-node regular solutions to Einstein-Yang-Mills-Higgs
theory for  quartic coupling $\lambda= 1/8$ and a range of
vacuum expectation values $v$.
Here and in figure (4), increasing $v$
corresponds to decreasing the
value of the radius at which the solution exponentially decays to its
vacuum value.
As $v$ decreases from its maximal value,
the quasi--$k=0$ solutions approach a flat space limit  which is the well known
Yang--Mills sphaleron. The other solutions do not admit a weak gravity limit.}

\FIG\fd{One- and two-node black hole solutions to Einstein-Yang-Mills-Higgs
theory for  quartic coupling $\lambda= 1/8$, a range of
vacuum expectation values $v$ and $r_h = 1$.  The quasi--$k=0$ solution
approaches a  Schwarzschild black hole metric as $v$ decreases and might be
heuristically described as a black hole with sphaleron hair. The other
solutions do not have a weak gravity limit (a limit arbitrarily close
to Schwarzschild).}

\ack
We would like to thank P. Argyres, T. Imbo, M. Ortiz, A. Shapere, D. Spector,
 H. Tye and S.T. Yau for helpful discussions.
This work was supported in part by the National Science Foundation, and by DOE
 grant DE-AC02-76ER 03069.

\vfil
\refout
\figout

\end